\documentstyle[12pt]{article}
\begin{document}
\begin{center}
\vspace{1cm}
  {\LARGE \bf  QCD EQUATIONS FOR GENERATING FUNCTIONALS \\
  AND MULTIPARTICLE CORRELATIONS }
\vspace{7mm}
{\large  I.M.\,DREMIN}
\vspace{5mm}
{\normalsize  Lebedev Physical Institute, Moscow 117924, Russia }
\end{center}
\begin{abstract}
QCD equations for generating functionals are solved at coinciding momenta of
particles. As a result, the relations for $q$-particle correlation functions
at equal momenta are obtained. They are directly connected with the previously
derived results about the factorial and cumulant moments of multiplicity
distributions. It is predicted that the correlations  at coinciding points
decrease, first, as a function of the rank $q$, and then start oscillating
when represented in a form of the ratio of cumulant to factorial correlations.
In particular, the cumulant function of the fifth rank should become negative
at the origin. Some experimental data are briefly discussed.
\end{abstract}

\section{INTRODUCTION}
Multiparticle correlations in hadroproduction at high energies are very
complicated both for experimental studies and for theoretical description.
Beside rather sophisticated dynamics, the problems due to the huge number
of available variables enter the game. The choice of any particular variable
leads to some specific correlations being emphasized (see, e.g., Ref.\cite{1}).

The multiparicle correlation functions integrated over momenta give rise to
the moments of multiplicity distributions. Some special features of the
moments have been revealed from the solution of QCD equations for generating
functions (for the review, see Ref.\cite{2}).

Surely, the dependence of the correlation functions on particle momenta
provides more detailed information. Unfortunately, their study in the
$3q$-dimensional phase space is not an easy task. That is why up to now
we have got experimental data on 2-particle (and, sometimes, on 3-particle)
correlation functions only plotted as functions of a particular variable
(usually, it is the relative (pseudo)rapidity or 4-momentum squared).

Theoretical approach asks for the equations for generating functionals.
In QCD they were written rather long time ago \cite{3,4}. However, no general
solution of them has been found. It is common to rewrite those equations in
the form of cascade equations {\em a'la} Altarelli-Parisi, and to find their
solution for inclusive distributions or for some specific correlations
\cite{3,4,5,6}.

We shall show how to get some information about correlations directly from
the equations for generating functionals using the experience we got from
the solution of corresponding equations for the generating functions \cite{2}.
We demonstrate it with the illustrative example of the relations between the
correlation functions of various ranks at coinciding momenta. In particular,
some predictions for the correlations of higher order (yet unavailable in
experiment) are obtained. The more sophisticated case of unequal momenta
will be considered in a separate publication.

\section{DEFINITIONS AND NOTATIONS}

The main ingredient of any correlation analysis is the $q$-particle inclusive
density
\begin{equation}
\rho _{q}^{(f_i)}({\bf k}_1,\ldots ,{\bf k}_q)\equiv \frac {1}{\sigma }\frac
{d^{q}\sigma ^{(f_i)}}{d\omega _1,\ldots ,d\omega _q} ,   \label{1}
\end{equation}
where the invariant momentum phase space is given by
\begin{equation}
d\omega _q = \frac {d^{3}{\bf k}_q}{(2\pi )^{3}2E_q} ,            \label{2}
\end{equation}
and the set of indices $f_i$ denotes the internal quantum numbers of all $q$
particles. We shall omit it in what follows.

The generating functional of inclusive distributions is defined by the formula
\begin{equation}
G[z(k)] = \sum _{q=0}^{\infty }\frac {1}{q!}\int d\omega _{1}\ldots \int
d\omega _{q} \rho _{q}z({\bf k}_1)\ldots z({\bf k}_q) ,   \label{3}
\end{equation}
where the integration is over available $3q$-dimensional phase space.

{}From (\ref{3}) one gets
\begin{equation}
\rho _{q}({\bf k}_1,\ldots ,{\bf k}_q) = \frac {\delta ^{q}G[z]}
        {\delta z({\bf k}_1)\ldots \delta z({\bf k}_q)} | _{z=0} .   \label{4}
\end{equation}
The normalized factorial correlation function is defined by
\begin{equation}
F_{q}({\bf k}_1,\ldots ,{\bf k}_q) = \frac {\rho _{q}({\bf k}_1,\ldots ,
{\bf k}_q)}{\rho _{1}({\bf k}_1)\ldots \rho _{1}({\bf k}_q)} .  \label{5}
\end{equation}

The logarithm of the functional (\ref{3}) is the generating functional of
the so-called cumulant correlation functions
\begin{equation}
\kappa _{q}({\bf k}_1,\ldots ,{\bf k}_q) = \frac {\delta ^{q}\ln G[z]}
{\delta z({\bf k}_1)\ldots \delta z({\bf k}_q)} | _{z=0} . \label{6}
\end{equation}
The corresponding normalized function is
\begin{equation}
K_{q}({\bf k}_1,\ldots ,{\bf k}_q) = \frac {\kappa _{q}({\bf k}_1,\ldots ,
{\bf k}_q)}{\rho _{1}({\bf k}_1)\ldots \rho _{1}({\bf k}_q)} . \label{7}
\end{equation}

In quantum field theory, the functions $\rho _q$ are determined by the whole
set of Feynman graphs while the functions $\kappa _q$ are given by the
connected graphs only. In correlation analysis, correspondingly, $\rho _q$
take into account all correlations, and $\kappa _q$ correspond to the
"genuine" $q$-particle correlations.

Integration of these functions over momenta gives rise to factorial and
cumulant moments of multiplicity distributions. Their generating function is
described by the same formula as (\ref{3}) at constant values of the
functions $z({\bf k})\equiv z =$ const.

\section{Equations for generating functionals and their solution}

The QCD Lagrangian defines the dynamics of hadroproduction, and, consequently,
the correlations of particles produced. It determines the relationship of
generating functionals in the form of non-linear integro-differential equations
which were proposed a long time ago \cite{3,4}. To simplify the equations, we
shall consider here the gluodynamics where quark degrees of freedom are
neglected. It is reasonable for qualitative conclusions we are aimed at as
our previous experience \cite{2} of solution of equations for generating
functions shows.

Thus the equation for the generating functional of gluodynamics is
\begin{equation}
\frac {\partial G(z(k),Y)}{\partial Y} = \int _{0}^{1}dxK(x)\gamma _{0}^{2}
[G(z,Y+\ln x)G(z,Y+\ln (1-x)) - G(z,Y)] .    \label{8}
\end{equation}
Here $Y=\ln (p\theta /p_{0}), p$ is the initial momentum, $\theta $ is the
angular width of the gluon jet considered, $p_0 =$const ,
\begin{equation}
\gamma _{0}^{2} = \frac {6\alpha _S}{\pi } ,  \label{9}
\end{equation}
$\alpha _S$ is the coupling constant, and the kernel of the equation is
\begin{equation}
K(x) = \frac {1}{x} - (1-x)[2-x(1-x)] .    \label{10}
\end{equation}
It is the non-linear integro-differential equation with shifted arguments in
the non-linear part which take into account the conservation laws, and with
the initial condition
\begin{equation}
G(z, Y=0) = 1+z(k=p) ,    \label{11}
\end{equation}
and the normalization
\begin{equation}
G(z=0, Y) = 1 .     \label{12}
\end{equation}
In particular, at constant $z$ one gets the generating function
\begin{equation}
G(z,Y) = \sum _{n=0}^{\infty }P_{n}(Y)(1+z)^n ,     \label{13}
\end{equation}
where $P_n$ is the probability of $n$-particle events.

The condition (\ref{13}) normalizes the total probability to 1, and the
condition (\ref{11}) declares that there is a single particle at the very
initial stage. The corresponding equation for the generating function
(\ref{13})
looks quite similar and is obtained from eq. (\ref{8}) at $z$ = const. Its
solution was considered in \cite{7}. In a similar way, one can get the
solution of (\ref{8}) at the special point where all the momenta of the
$q$ registered particles (actually, gluons in our treatment) are equal i.e.
${\bf k}_1 = {\bf k}_2 = \ldots ={\bf k}_q$. Therefore, the momenta differences
are chosen equal to zero. It means that we are considering the correlations at
coinciding points where $\Delta {\bf k} = 0$. Since the correlation functions
depend on the differences of momenta and on the overall momentum (i.e. on $Y$)
the only dependence left in our case is the dependence on $Y$.

After Taylor series expansion and differentiation in eq. (\ref{8}) similar to
the procedure in \cite{7}, one gets the same differential equation
\begin{equation}
(\ln G(Y))\prime \prime = \gamma _{0}^{2}[G(Y)-1-2h_{1}G\prime (Y)+
h_{2}G\prime \prime (Y)] ,    \label{14}
\end{equation}
where $h_1 = 11/24; h_2 = (67-6\pi ^2)/36\approx 0.216$.

In general, one can define the anomalous dimension $\gamma
(\omega _i, \alpha _S(Y))$ by the formula which generalizes ones commonly
used for average multiplicity or single inclusive distributions (see
\cite{3,4})
\begin{equation}
\rho _q(\omega _i,Y) = \rho _q(Y_0)\exp [q\int _{Y_0}^{Y}dy\gamma (\omega _i,
\alpha _S(y))] .   \label{15}
\end{equation}

Substituting (\ref{15}) in (\ref{14}) one gets after differentiation for terms
with equal powers of $z(\omega _i)$
\begin{eqnarray}
H_q(\omega _i) = \frac {K_q(\omega _i)}{F_q(\omega _i)} = \frac {\kappa _q(
\omega _i)}{\rho _q(\omega _i)} = \nonumber \\ \frac {\gamma _{0}^{2}[1-2h_{1}q
\gamma (\omega _i)+h_{2}(q^2\gamma ^2(\omega _i)+q\gamma \prime (\omega _i)]}
{q^2\gamma ^2(\omega _i)+q\gamma \prime (\omega _i)} . \label{16}
\end{eqnarray}
Even though we write down the explicit dependence on $\omega _i$, it was
assumed
in (\ref{14}) when differentiating over $Y$ that correlation functions are
constant in $\omega _i$ i.e. the correlation lengths are much larger than
the interval of momenta considered. That is why we omit $\omega _i$ in what
follows considering it equal to zero. Apart from it, the condition $qk\ll E$,
where $E$ is the total energy,
should be fulfilled to ensure  no additional dependence on $Y$. Then all
calculations are just the same as for the moments of multiplicity
distributions in Ref.\cite{7}, and from the normalization condition
$H_1(0)=1$ one gets the same relation between $\gamma (0)$ and $\gamma _0$
\begin{equation}
\gamma (0)\approx \gamma _{0}-\frac {1}{2}h_{1}\gamma _{0}^{2}+\frac {1}{8}
(4h_{2}-h_{1}^{2})\gamma _{0}^{3}+O(\gamma _{0}^{4}) .   \label{17}
\end{equation}
The value of $\gamma _0$ is measured in experiment (it is equal 0.48 at
the mass of $Z^0$-boson). Therefore it is possible to estimate the values
of correlation functions at coinciding points according to (\ref{16}) and
(\ref{17}).

The main conclusion is that the ratio of cumulant to factorial correlation
functions at the origin should be approximately equal to the ratio of their
integrals
\begin{equation}
H_q(0)=\frac {K_q(0)}{F_q(0)}\approx \frac {K_q}{F_q}=H_q .   \label{18}
\end{equation}
According to results of \cite{7,2}, it predicts that this ratio should
decrease fastly at $q=2, 3, 4$ and reach the negative value at $q=5$.
At ever higher values of $q$ it should oscillate (for the review see \cite{2}).

\section{Theoretical results and experimental data}

Our main theoretical statement relates two sets of experimentally accessible
quantities, namely, moments of multiplicity distributions and the values of
correlation functions at the origin. It could be verified without any
theoretical formulae. However, one should clearly recognize possible
shortcomings which could be of various origin. First, the prediction is done
for gluons while in experiment we deal with pions mostly. However, the
success in predicting the corresponding qualitative
features of multiplicity distributions (see \cite{2}) is
encouraging. More important problems could be related to hadronization effects
just at the origin because of resonances, Bose-Einstein correlations etc.
Moreover, we have now the experimental data only about two- and three-particle
correlation functions (with rather low precision) while the data about
moments of multiplicity distributions are much more precise and admit good
predictions. It appeals to experimentalists.

At the moment I was able to get the necessary ratios from the data about
densities $\rho _q$ provided by NA22 Collaboration using the formulae
\begin{equation}
H_2(0) = \frac {C_2(0)}{\rho _2(0)} = 1-\frac {\rho _{1}^{2}(0)}{\rho _{2}(0)}
,
\label{19}
\end{equation}
\begin{equation}
H_3(0) = \frac {C_3(0)}{\rho _3(0)} =\frac {\rho _3(0)-3\rho _1(0)\rho _2(0)+
2\rho _{1}^{3}(0)}{\rho _{3}(0)} .   \label{20}
\end{equation}
Using the data for non-singlediffractive $\pi ^{+}p$ events at 250 GeV/c, one
gets
\begin{equation}
H_{2}^{cc}(0)=0.34 ; H_{3}^{ccc}(0)=0.13 ; \label{21}
\end{equation}
\begin{equation}
H_{2}^{--~--}(0)=0.28 ; H_{3}^{--~--~--}(0)=0.1 ;     \label{22}
\end{equation}
\begin{equation}
H_{2}^{++}(0)=0.25 ; H_{3}^{+++}(0)=0.07   \label{23}
\end{equation}
for charged $(c)$, negatives $(--)$ and positives $(+)$.
These values agree qualitatively with those shown in \cite{2} for
moments of multiplicity distributions in $p\bar p$
data (e.g., UA5 Collaboration data for charged-charged correlations are
$H_2 = 0.26 ; H_3 = 0.07$) and show the same tendency as the ratios of
moments. Hardly, more elaborated tests can be done at present stage but one
could try to check it in Monte Carlo models as well. The most impressive
prediction of the negative value of the fifth order correlation function
at the origin can be also verified in Monte Carlo models.

\section{Conclusions}

The solution of the equations for generating functionals in gluodynamics
gives a hint to the equality of the ratios of integral cumulant and factorial
moments to the corresponding ratios of the values of cumulant and
factorial correlation functions at the origin. In particular, it follows
that latest ratio should also decrease from $q=1$ to $q=5$ becoming negative,
and
oscillating at ever higher values of ranks $q$. It would be desirable to
verify the prediction using the data of same experiments. The comparison
to Monte Carlo models is welcome too. From the theoretical side it would
indicate once again the importance of a parameter $q\gamma $ in QCD
(at the inclusive level, it is closely related to the exclusive factor
$n!\alpha _S^n$ widely discussed in
vector boson production at threshold and in connection with Parke-Taylor
amplitudes). Besides, the negative values of the ratio $H_5$
at asymptotically high energies would show
the decline from the negative binomial distribution which predicts the
positive values of $H_q$ at any $q$.

\vspace{5mm}
{\large ACKNOWLEDGEMENTS}
\vspace{2mm}
This work is supported by the Russian Fund for Fundamental Research (grant
93-02-3815), by Soros Foundation (grant M5V000) and by the JSPS Fund.\\
My special thanks are to F.\,Rizatdinova for providing experimental data on
various species densities.

\end{document}